\documentclass[dvipdfmx]{pasj01}
\draft
\begin{document}

\title{{Lithium} in CEMP-no stars: A new constraint on the lithium depletion mechanism in the early universe\footnotemark{$^\dagger$}}
\author{Tadafumi \textsc{Matsuno}\altaffilmark{1,2}}
\altaffiltext{1}{Department of Astronomical Science, School of Physical Sciences, SOKENDAI (The Graduate University for Advanced Studies), Mitaka, Tokyo 181-8588, Japan}
\altaffiltext{2}{National Astronomical Observatory of Japan (NAOJ), 2-21-1 Osawa, Mitaka, Tokyo 181-8588, Japan}
\email{tadafumi.matsuno@nao.ac.jp}
\author{Wako \textsc{Aoki}\altaffilmark{1,2}}
\author{Takuma \textsc{Suda}\altaffilmark{3}}
\altaffiltext{3}{Research Center for the Early Universe, University of Tokyo, 7-3-1 Hongo, Bunkyo-ku, Tokyo 113-0033, Japan}
\author{Haining \textsc{Li}\altaffilmark{4}}
\altaffiltext{4}{Key Lab of Optical Astronomy, National Astronomical Observatories, Chinese Academy of Science, A20 Datun Road, Chaoyang, Beijing 100012, China}
\KeyWords{Stars: Population II --- Stars: abundances --- early universe}
\maketitle
\footnotetext[$^\dagger$]{Based on data collected with the Subaru Telescope, which is operated by the National Astronomical Observatory of Japan}
\begin{abstract}

Most of relatively warm, unevolved metal-poor stars ($T_{\rm eff}\gtrsim 5800\,\mathrm{K}$ and $[{\rm Fe/H}]\lesssim -1.5$) exhibit almost constant lithium abundances, irrespective of metallicity or effective temperature, and thus form the so-called Spite plateau. 
This was originally interpreted as arising from lithium created by the Big Bang nucleosynthesis. 
Recent observations, however, have revealed that ultra metal-poor stars (UMP stars; $[\mathrm{Fe/H}]<-4.0$) have significantly lower lithium abundances than that of the plateau. 
Since most of the UMP stars are CEMP-no stars, carbon-enhanced metal-poor stars with no excess of neutron-capture elements, a connection between the carbon enhancement and lithium depletion is suspected.
A straightforward approach to this question is to investigate carbon-normal UMP stars.
However only one object is known in this class. 
As an alternative, we have determined lithium abundances for two CEMP-no main-sequence turn-off stars with metallicities $[{\rm Fe/H}]\sim -3.0$, where there are numerous carbon-normal stars with available lithium abundances that can be considered. 
Our 1D LTE analysis indicates that the two CEMP-no stars have lithium abundances that are consistent with values near the plateau, which suggests that carbon enhancement and lithium depletion are not directly related.  
Instead, our results suggest that extremely low iron abundance is a fundamental cause to depleted lithium in UMP stars.

\end{abstract}

\section{Introduction}\label{secint}

Very metal-poor stars preserve the chemical composition of their natal gas clouds, hence their elemental abundances have been extensively used to investigate the nature of the first nucleosynthesis processes in the Universe (see, e.g., \cite{Beers2005}; \cite{Frebel2015}). 
Lithium is of particular interest, because it is the only ``metal'' created during the Big Bang. 
\citet{Spite1982} first pointed out that metal-poor dwarfs ($T_{\rm eff} \gtrsim 5800\, \mathrm{K}$ and $[\mathrm{Fe/H}]\lesssim -1.5$\footnote{$[\mathrm{X/Y}]= \log(N_{\rm X}/N_{\rm Y})-\log(N_{\rm X}/N_{\rm Y})_\odot$}) have an almost constant lithium abundance, regardless of their effective temperature or metallicity (the so-called ``Spite plateau''). 
The temperature range is limited because lithium abundances dramatically decrease in cooler stars due to deep convection during stellar evolution.
In the metallicity range of $[\mathrm{Fe/H}]>-1.5$, production of lithium via, for instance, novae and cosmic ray, can affect the lithium abundances in more metal-rich stars (see \cite{Prantzos2012} for more details).
In any event, until relatively recently, the constant value of $A(\rm Li)=2.2\pm0.1$\footnote{$A({\rm X}) =\log (N_{\rm X}/N_{\rm H}) +12$} among unevolved metal-poor stars (e.g., Spite \& Spite 1982) was considered likely to represent the primordial lithium abundance emerging from the Big Bang nucleosynthesis.

Doubts concerning this interpretation emerged when increasingly precise inferences of the expected primordial lithium abundance began to be made, in particular those based on derived estimates of the baryon number density from cosmic microwave background measurements combined with standard Big Bang nucleosynthesis models (\cite{Coc2004}; \cite{Spergel2007}). 
In the modern era of precision cosmology, \citet{Coc2014} derived the predicted primordial value of lithium to be $A(\rm Li) =2.66-2.73$ based on well-constrained cosmological parameters \citep{PlanckCollaboration2014}. 
This value is higher than the observed value of the lithium abundances for stars on the Spite plateau by $\sim 0.5 \,\mathrm{dex}$. 

Following the early studies, observations of metal-poor stars with lower metallicity, which is expected to approach more closely the primordial lithium abundance, have highlighted additional difficulties. 
\citet{Ryan1999} pointed out that there is a slope of the plateau at [Fe/H]$<-2.5$ though the star-to-star scatter is very small. 
They concluded that lithium in metal-poor stars is not primordial but affected by some processes other than the Big Bang, in particular the Galactic chemical evolution.
More recent studies have reported the so-called ``breakdown'' of the Spite plateau at extremely low metallicity, e.g., $[{\rm Fe/H}]\lesssim -3$, where measured lithium abundances start to exhibit scatter, and extend the mean value to even lower levels (e.g., \cite{Aoki2009}; \cite{Sbordone2010}). 
More significantly, no ultra metal-poor (UMP; [Fe/H]$<-4$) star has been reported to be on the plateau (e.g., \cite{Bonifacio2015}). 
Although the convection mechanisms which can lead to depletion of lithium are expected to be less effective at lower metallicity, some authors have invoked alternative stellar processes capable of depleting lithium that might be active in very metal-poor stars \citep{Melendez2010}.
If such a depletion mechanism inside UMP stars themselves does work, similar mechanism could bring lithium abundances in less metal-poor stars from $A({\rm Li})\simeq 2.7$ to the Spite plateau value.

A remarkable feature of UMP stars is the high fraction of carbon enhanced objects. 
All of carbon-enhanced UMP stars show no-excess of heavy neutron-capture elements \citep{Norris2012}, and are classified into so-called CEMP-no stars (Carbon Enhanced Metal-Poor stars without heavy elements enhancement; \cite{Beers2005}). 
For example, HE 1327-2326, which is the most metal-poor main sequence turn-off star with enhanced carbon abundance, has a very low upper limit on the lithium abundance, $A({\rm Li})<0.70$ (\cite{Frebel2005}; \cite{Aoki2006}; \cite{Frebel2008}).  
The fact that most of UMP stars are CEMP-no stars leads us to investigate the connection between lithium depletion and carbon excess. 

From theoretical view points, a possible relation between carbon enhancement and lithium depletion has been proposed. 
\citet{Piau2006} suggested that lithium astration in population III stars, in addition to lithium destruction in stars which are now observed as metal-poor stars, is a possible solution to the lithium problems. 
The large enhancement of carbon in UMP stars could be attributed to unique yields from population III stars (e.g., spin star model; \cite{Maeder2015}, faint supernovae model; \cite{Nomoto2013}, \cite{Tominaga2014}). 
Hence, it is qualitatively possible to assume that UMP stars are formed under strong influence of population III stars, which results in carbon enhancement and lithium depletion. 
Another explanation is that CEMP-no stars underwent mass transfer from former AGB companion in which carbon is enhanced \citep{Suda2004}. 
The same mechanism is considered to be responsible for the formation of CEMP-s stars (CEMP stars with s-process elements enhancement). 
Since CEMP-s stars are not yet found among UMP stars, CEMP-no stars might be a low metallicity counterpart of CEMP-s stars. 
{Note that reported binary frequency of CEMP-s stars is $\sim 82\pm 10\%$ whereas that of CEMP-no stars is $17\pm 9\%$ (\cite{Hansen2016b}; \cite{Hansen2016}).
While lithium might be produced at some phase of AGB evolution through the Cameron-Fowler mechanism \citep{Sackmann1992}, lithium is more generally destroyed. 
In fact, lithium abundance in CEMP-s stars is usually lower depending on the amount of the accreted mass \citep{Masseron2012}. 

In order to examine the possible connection between lithium depletion and carbon excess at very low metallicity, it is highly desirable to observe carbon-normal stars at $[{\rm Fe / H}]<-4$. 
However, only one unevolved carbon-normal UMP star has been reported (SDSS J102915+172927; \cite{Caffau2012}), in spite of recent extensive surveys and their follow-up high resolution spectroscopic observations (e.g., \cite{Aoki2013}; \cite{Roederer2014}). 
Interestingly, this star has very low lithium abundance ($A({\rm Li})<1.1$), whereas the upper limit of the carbon abundance is low ($[{\rm C/Fe}]<0.93$).   

An alternative approach is to investigate lithium abundances in CEMP-no stars with $[\mathrm{Fe/H}]\sim -3$ with the assumption that they have the same origin as CEMP-no stars with $[\mathrm{Fe/H}]<-4$. \citet{Masseron2012} reported low lithium abundances in CEMP stars without classifying them into subclasses. 
However, each subclass of CEMP stars should be discussed separately since they could be formed through different formation processes. 
The current sample is not large enough to discuss the lithium abundance of CEMP-no stars statistically. 
Moreover, in order to discuss the relation between lithium and carbon, we need to compare lithium abundances in CEMP-no stars with those of carbon-normal stars with similar metallicity. 

In this paper, we determine the lithium abundances in two unevolved CEMP-no stars as well as in comparison stars with normal carbon abundances with $[{\rm Fe / H}]\sim-3$. 
In Section \ref{secobs}, we briefly explain how these objects have been selected and observed. 
Methods to estimate stellar parameters and to measure abundances are described along with the results in Section \ref{secmeth}. 
Finally we give interpretation to them in Section \ref{secdis}.

\section{Target selection \& Observation}\label{secobs}

The sample of the present abundance study includes two CEMP-no stars. 
One of them, SDSS J1424+5615, is selected from the extremely metal-poor stars studied by \citet{Aoki2013}, who derived $T_{\rm eff}=6350\,\mathrm{K}$ and $[\mathrm{Fe/H}]=-2.97$ for this object by the ``snap-shot'' spectroscopy.
A high resolution spectrum with $R=60,000$ covering $4030-6800\,\mathrm{\AA}$ was obtained with the Subaru Telescope High Dispersion Spectrograph (HDS; \cite{Noguchi2002}) on June 29, 2009 (see Table 1). 
Stellar parameters and elemental abundances discussed in this paper are determined from this new spectrum.

The other CEMP-no star, LAMOST J1410-0555, is identified in the follow-up high-resolution spectroscopy with the Subaru Telescope for candidate metal-poor stars found in the Data Release 1 of LAMOST (the Large sky Area Multi-Object fiber Spectroscopic Telescope; \cite{Cui2012}; \cite{Luo2012}; \cite{Zhao2012}; \cite{Luo2015}). 
A high resolution spectrum with $R=45,000$ for $4030-6800\,\mathrm{\AA}$ was obtained on May 10, 2014. 
The spectrum of a comparison star, LAMOST J1305+2815, was obtained with the same setup of HDS (see Table \ref{tabobs}). 
See \citet{Li2015a} for more details on the metal-poor star survey with LAMOST and follow-up study with the Subaru Telescope.

Standard data reduction was conducted using the IRAF echelle package 
\footnote{IRAF is distributed by the National Optical Astronomy Observatory, which is operated by the Association of Universities for Research in Astronomy, Inc. under cooperative agreement with the National Science Foundation.}
, including bias level correction for the CCD data, scattered light subtraction, flat-fielding, extraction of spectra, and wavelength calibration using Th arc lines. 
The signal-to-noise (S/N) ratios per $1.8\,\mathrm{km\,s^{-1}}$ pixel are estimated from the standard deviation of photon counts for continuum at $6708\,\mathrm{\AA}$ (Table \ref{tabobs}).
Heliocentric radial velocities are also provided in Table \ref{tabobs}. 
We also analyze the spectrum of G~64-12 obtained by \citet{Aoki2006} for comparison purposes. 
This is a bright extremely metal-poor ($[\mathrm{Fe/H}]\simeq -3.3$) main-sequence turn-off star subject to many previous studies (e.g., \cite{Aoki2009}; \cite{Aoki2006}; \cite{Barklem2005}; {Ishigaki} et~al. 2012; \cite{Stephens2002}; {Ishigaki} et~al. 2010; \cite{Roederer2014}; \cite{Placco2016}).

\begin{table*}[tbph]
  \tbl{Observations for our targets.}{
  \begin{tabular}{lrrrrrr}\hline
  Obs.Name          & R.A.        & DEC .       & Obs.Date      & Exp.Time & S/N                    & $v_r$ \\
                    &             &             &               & (min)    & ($6700\,\mathrm{\AA}$) & ($\mathrm{km\,s^{-1}}$) \\ \hline
  LAMOST J1410-0555 & 14 10 02.83 & -05 55 51.8 & May 10, 2014  & 15       & 113                    & 94.89\\
  SDSS   J1424+5615 & 14 24 41.88 & +56 15 35.0 & June 29, 2009 & 230      & 95                     & -0.13\\
  LAMOST J1305+2815 & 13 05 34.80 & +28 15 10.5 & May 11, 2014  & 15       & 88                     & 35.84\\\hline
  \end{tabular}}
  \label{tabobs}
\end{table*}

\section{Abundance analysis \& Results}\label{secmeth}
We apply the standard abundance analysis technique using 1D model atmospheres of the ATLAS NEWODF grid assuming no convective overshooting \citep{Castelli2003}. 
$\alpha$-enhanced models ($[\alpha/\mathrm{Fe}]=0.4$) are used since our targets show $\alpha$ elements enhancement.
Our results are compared with previous ones for a large number of carbon-rich / carbon-normal stars analysed by 1D models. 
Though 3D effect is beyond the scope of this paper, it should be noted that \citet{Sbordone2010} and \citet{Lind2009} reported that 3D and NLTE correction required for lithium abundances is systematic and not larger than $0.05\,\mathrm{dex}$. 

\subsection{Stellar parameters}
We carried out analysis of Balmer line profiles in order to estimate effective temperature and surface gravity. 
Model profiles of Balmer lines are computed by \citet{Barklem2002}\footnote{http://www.astro.uu.se/\%7ebarklem/}. 
They calculated them with 1D LTE model atmospheres of MARCS of the epoch of 1997 (\cite{Asplund1997} and references therein), using self broadening theory by \citet{Barklem2000}. 
Assumed parameters of mixing-length theory (MLT) are $\alpha=0.5$ and $y=0.5$. 
The grid of the model profiles covers effective temperatures from $4500\,\mathrm{K}$ to $7500\,\mathrm{K}$ with interval of $100\,\mathrm{K}$. 
Surface gravity of the grid spans from $1.0$ to $5.0$ for $T_{\rm eff} \leq 5500\,\mathrm{K}$ and $3.4$ to $5.0$ for $T_{\rm eff} \geq 5600\,\mathrm{K}$. 
For each set of $T_{\rm eff}$ and $\log g$, metallicity spans from $[{\rm Fe / H}]=-3.0$ to $[{\rm Fe / H}]=1.0$ with interval of $0.5$. 
$[\alpha/{\rm Fe}]=0.4$ models are used during Balmer lines analysis.

Normalization of observed spectra around a broad Balmer line is essential to estimate stellar parameters. 
Following the method described in \citet{Barklem2002}, multi-order spectra that do not include Balmer lines are normalized by fitting continuum level with spline function, and the continuum profile of the orders including a Balmer line is estimated by interpolation from adjacent two orders.
 
Stellar parameters are adopted as model profile fitting gives the minimum $\chi ^2$ for the shaded regions in Figure \ref{figbalmer1}. 
Firstly, we make an estimation of $T_{\rm eff}$ from the wings of H$\beta$ line assuming $\log g=4.0$ since the shape of the line is sensitive to $T_{\rm eff}$ but not to $\log g$ or metallicity. 
With the estimated $T_{\rm eff}$, $\log g$ is subsequently estimated from the wings of H$\alpha$, which depend on both of $\log g$ and $T_{\rm eff}$. 
Then, $T_{\rm eff}$ is redetermined from H$\beta$ with the estimated $\log g$. 
Finally, $\log g$ is redetermined from H$\alpha$ with the updated $T_{\rm eff}$. 
The fitting results of Balmer lines for LAMOST J1410-0555 are shown in Figure \ref{figbalmer1}. 
Metallicity is fixed to [Fe/H]$=-3$ through Balmer line analysis, though metallicity dependence on derived parameters is small.

Stellar parameters obtained for the observed stars together with G64-12 are given in Table \ref{tabab} and plotted in Figure \ref{figHR}. 
All of our program stars are in the main sequence or subgiant phase, i.e., before the first dredge-up.
\begin{figure*}[htbp]
  \begin{center}
  \FigureFile(160mm,112mm){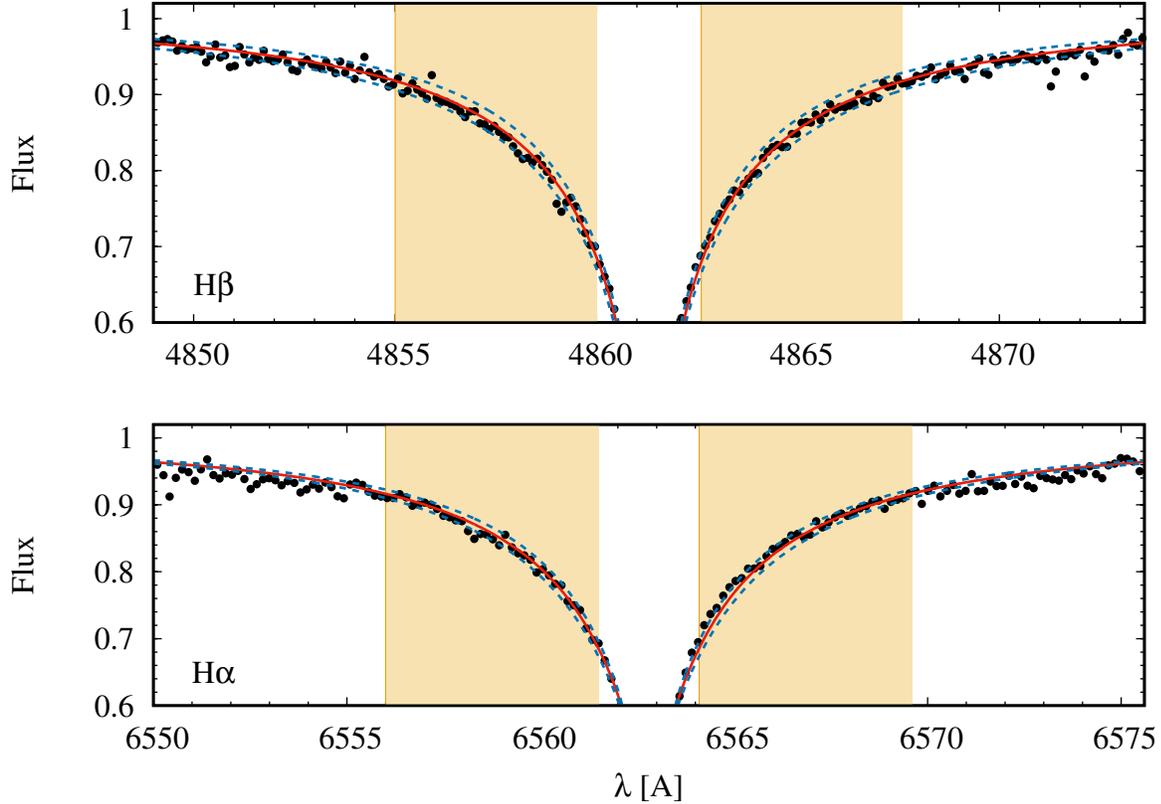}
  \end{center}
  \caption{Fitting results for H$\beta$ and H$\alpha$ of LAMOST J1410-0555. Red solid lines show the best fit model spectrum, $T_{\rm eff}=6169\,\mathrm{K}$ and $\log g =4.21$. Blue dashed lines show model spectra with $\pm 0.3\,\mathrm{dex}$ offset in $\log g$ for H$\alpha$, and $\pm 100\,\mathrm{K}$ offset in $T_{\rm eff}$ for H$\beta$. Yellow shaded regions are used in fitting.}
  \label{figbalmer1}
\end{figure*}
\begin{table*}
  \tbl{Parameters and abundances of our targets.}{
{\tabcolsep = 1.2mm
  \begin{tabular}{llcrrrrcrrrrcrrrrcrrrrc}\hline
  &&&\multicolumn{4}{c}{LAMOST J1410-0555}& &\multicolumn{4}{c}{SDSS J1424+5615}&&\multicolumn{4}{c}{LAMOST J1305+2815}&&\multicolumn{4}{c}{G64-12}&\\\cline{4-7}\cline{9-12}\cline{14-17}\cline{19-22}
                                        &       &  &       & $\sigma $ & $(N)$ & error &  &          & $\sigma $ & $(N)$ & error &  &         & $\sigma $ & $(N)$ & error &  &         & $\sigma $ & $(N)$ & error & \\ \hline
                     $T_{\rm eff}$      &       &  & 6169  &           &       & 40    &  & 6088     &           &       & 40    &  & 6102    &           &       & 40    &  & 6291    &           &       & 40    & \\
                     $\log g$           &       &  & 4.21  &           &       & 0.2   &  & 4.34     &           &       & 0.2   &  & 3.75    &           &       & 0.2   &  & 4.38    &           &       & 0.2   & \\
                     $v_{\rm turb}$     &       &  & 1.1   &           &       &0.3   &  & 1.1      &           &       & 0.3   &  & 1.5     &           &       & 0.3   &  & 1.5     &           &       & 0.3   & \\
                     $[\mathrm{Fe/H}]$  & Fe I  &  & -3.19 & 0.11      & 38    & 0.04  &  & -3.01    & 0.10      & 31    & 0.04  &  & -2.93   & 0.09      & 51    & 0.04  &  & -3.36   & 0.05      & 49    & 0.03  & \\ \hline
                     $A$(Li)            & Li I  &  & 2.17  & ----      & s     & 0.12  &  & 2.14     & ----      & s     & 0.11  &  & 2.10    & ----      & s     & 0.10  &  & 2.22    & ----      & s     & 0.06  & \\
                     $[\mathrm{C/Fe}]$  & CH    &  & 1.53  & ----      & s     & 0.12  &  & 1.49     & ----      & s     & 0.11  &  & $<$1.30 & ----      & s     & 0.11  &  & $<$0.93 & ----      & s     & 0.08  & \\
                     $[\mathrm{Na/Fe}]$ & Na I  &  & 0.56  & 0.05      & 2     & 0.08  &  & 0.40     & 0.04      & 2     & 0.08  &  & -0.07   & 0.06      & 2     & 0.07  &  & -0.09   & 0.01      & 2     & 0.05  & \\
                     $[\mathrm{Mg/Fe}]$ & Mg I  &  & 0.84  & 0.08      & 6     & 0.04  &  & 0.74     & 0.07      & 6     & 0.04  &  & 0.30    & 0.08      & 7     & 0.04  &  & 0.36    & 0.04      & 5     & 0.03  & \\
                     $[\mathrm{Ca/Fe}]$ & Ca I  &  & 0.34  & 0.09      & 10    & 0.03  &  & 0.16     & 0.17      & 6     & 0.07  &  & 0.43    & 0.08      & 16    & 0.02  &  & 0.51    & 0.06      & 18    & 0.02  & \\
                     $[\mathrm{Sc/Fe}]$ & Sc II &  & 0.26  & 0.07      & 2     & 0.10  &  & 0.35     & 0.09      & 2     & 0.10  &  & 0.34    & 0.03      & 7     & 0.07  &  & 0.28    & 0.12      & 4     & 0.09  & \\
                     $[\mathrm{Ti/Fe}]$ & Ti I  &  & 0.52  & 0.06      & 3     & 0.06  &  & 0.83     & 0.00      & 1     & 0.10  &  & 0.64    & 0.11      & 5     & 0.05  &  & 0.66    & 0.09      & 6     & 0.04  & \\
                     $[\mathrm{Ti/Fe}]$ & Ti II &  & 0.23  & 0.06      & 10    & 0.07  &  & 0.28     & 0.07      & 9     & 0.07  &  & 0.48    & 0.10      & 17    & 0.07  &  & 0.57    & 0.05      & 18    & 0.07  & \\
                     $[\mathrm{Cr/Fe}]$ & Cr I  &  & -0.32 & 0.05      & 5     & 0.02  &  & -0.30    & 0.11      & 4     & 0.06  &  & -0.22   & 0.04      & 5     & 0.02  &  & -0.11   & 0.02      & 4     & 0.01  & \\
                     $[\mathrm{Fe/Fe}]$ & Fe II &  & -0.03 & 0.12      & 6     & 0.09  &  & 0.11     & 0.08      & 6     & 0.08  &  & -0.04   & 0.11      & 11    & 0.08  &  & 0.11    & 0.07      & 12    & 0.08  & \\
                     $[\mathrm{Sr/Fe}]$ & Sr II &  & -0.10 & 0.00      & 1     & 0.12  &  & -0.40    & 0.00      & 1     & 0.11  &  & 0.28    & 0.00      & 1     & 0.10  &  &         & ----      & 0     & ----  & \\
                     $[\mathrm{Ba/Fe}]$ & Ba II &  & -0.33 & 0.00      & 2     & 0.10  &  & $<$-0.69 & ----      & --    & ----  &  & -0.38   & 0.05      & 3     & 0.09  &  & -0.08   & 0.02      & 2     & 0.08  & \\\hline
\end{tabular}}}
  \label{tabab}
  \begin{tabnote}
  Fe in $[\mathrm{X/Fe}]$ denotes iron abundance from Fe~I lines. Solar abundances are from \citet{Asplund2009}. Total error for each abundance includes $\sigma^2/N$ and those due to internal error in stellar parameters.
  \end{tabnote}
\end{table*}

In most cases, $T_{\rm eff}$ and $\log g$ from Balmer lines yield consistent results with ionization and excitation balance of iron. 
$v_{\rm turb}$ is determined not to produce a trend in iron abundance with equivalent width (see Table \ref{tabab}). 
The uncertainty of $v_{\rm turb}$ is determined so that the trend does not become significant at $1\sigma$ level.

Uncertainties of $T_{\rm eff}$ and $\log g$ due to Poisson noise are estimated using a Monte-Carlo method. 
A hundred spectra are generated by adding noise to a synthetic spectrum calculated by the model atmosphere of $T_{\rm eff}=6200\,\mathrm{K}$, $\log g=4.2$ and $[\mathrm{Fe/H}]=-3.0$ assuming the noise having a Gaussian distribution of which standard deviation is $(\mathrm{S/N})^{-1}$. 
$T_{\rm eff}$ and $\log g$ are re-determined for these spectra with above procedure of profile fitting for Balmer lines. 
We take standard deviation of the results obtained for the 100 spectra as uncertainty. 
The resultant uncertainty is $\sim 5\,\mathrm{K}$ in $T_{\rm eff}$ and $\sim 0.02\,\mathrm{dex}$ in $\log g$ for a spectrum with $\mathrm{S/N}\sim 100$, indicating that the uncertainty due to random noise of spectra is sufficiently small. 
On the other hand, uncertainty of continuum level has larger impact on stellar parameters. 
The accuracy of interpolation in the normalization procedure is estimated to be $0.5\,\%$ by comparing continuum spectra from interpolation with those from direct normalization technique for orders those are free from broad absorption lines. 
We assume that the accuracy of interpolation does not significantly depend on S/N for spectra with S/N$>50$ and with [Fe/H]$<-3$, where the number of absorption lines is small. 
$0.5\,\%$ change in continuum level leads to $\Delta T_{\rm eff}\sim 35\,\mathrm{K}$ and $\Delta \log g \sim 0.15\,\mathrm{dex}$. 
Therefore, we take $\Delta T_{\rm eff}\sim40\,\mathrm{K}$ and $\Delta \log g\sim0.2\,\mathrm{dex}$ as total uncertainty for stellar parameters. 

In addition to these random errors, systematic errors due to uncertainty of model profile calculations should exist. 
A possible estimate of the systematic errors is to compare the stellar parameters with values obtained by other methods. 
\citet{Norris2013} determined effective temperature and surface gravity for metal-poor stars through three different procedures: spectrophotometry, Balmer line profile fitting and H$\delta$ index HP2. 
They reported systematically lower effective temperature from Balmer lines than that from spectrophotometry by $128\,\mathrm{K}$ in average. 
We note that Balmer line profiles computed from 3D hydrodynamical calculation of stellar atmosphere are sharper than those from 1D calculation assuming $\alpha=0.5$, which leads to hotter $T_{\rm eff}$ in 3D Balmer line analysis \citep{Ludwig2009}. 
Adopting a temperature lower by $128\,\mathrm{K}$ at $T_{\rm eff}\sim 6200\,\mathrm{K}$ leads to $0.10\,\mathrm{dex}$ lower $[\mathrm{Fe/H}]$, $0.09\,\mathrm{dex}$ lower $A({\rm Li})$, $0.12\,\mathrm{dex}$ lower $[\mathrm{C/Fe}]$ and $0.02\,\mathrm{dex}$ higher $[\mathrm{Ba/Fe}]$. 
 
Very recently, the data release 1 of the Gaia satellite \citep{GaiaCollaboration2016} became available.
The data release contains parallaxes of LAMOST J1410-0555, LAMOST 1305+2815 and G64-12.
We derive $\log g$ for them from their parallaxes and photometric data (APASS DR9 for LAMOST J1410-0555 and G64-12, \cite{Henden2016}; SDSS DR10 for LAMOST J1305+2815, \cite{Ahn2014}).
Computed $\log g$ are $4.3\pm 0.1$ for LAMOST J1410-0555, $4.3\pm0.1$ for G64-12, and $3.9^{+0.3}_{-0.7}$ for LAMOST J1305+2815. Our $\log g$ from H$\alpha$ lines are in fairly good agreement with those derived from the Gaia parallaxes. 

We here compare stellar parameters of some individual stars with those from previous studies. 
Stellar parameters of G64-12 from various studies are listed in Table \ref{tabg6412}. 
Our $T_{\rm eff}$ estimation is consistent with previous ones obtained by Balmer line analysis, whereas $T_{\rm eff}$ from excitation equilibrium of iron is systematically lower (\cite{Stephens2002}; {Ishigaki} et~al. 2010; \cite{Roederer2014}). 
Similar trend has been reported for very metal-poor giants by \citet{Frebel2013}.
The three studies that derived lower surface gravity ($\log g<4$), are based on excitation equilibrium of iron to derive effective temperature (Table \ref{tabg6412}). 
The discrepancy in $\log g$ between our study and Stephens and Boesgaard (2002) or \citet{Ishigaki2010} is partly due to lower effective temperature. 
They determined $\log g$ by requiring ionization equilibrium. 
In order to compensate neutralization by lower $T_{\rm eff}$, $\log g$ has to be lowered by $\sim0.2\,\mathrm{dex}$ per $100\,\mathrm{K}$. 
In addition, non-LTE (NLTE) effect should cause overionization, and affect iron abundance derived from Fe~I lines whereas that from Fe~II lines is not changed (e.g., \cite{Thevenin1999}). 
Therefore, lower $\log g$ is needed to achieve ionization equilibrium in 1D LTE analysis since it derives higher iron abundance from Fe~II lines than that from Fe~I lines. 
Likely these two processes cause the discrepancy.
On the other hand, a lower $\log g$ is obtained by \citet{Roederer2014} assuming its evolutionary phase from isochrones of subgiant.
We note that the Gaia parallax indicates that G64-12 is in the main sequence phase.

\citet{Aoki2013} derived $T_{\rm eff}$ of SDSS J1424+5615 as 5971 K from its $(V-K)$ color and 6266 K from its $(g-r)$ color, but adopted 6339 K based on the estimate by the SDSS SEGUE pipeline (SSPP) of DR7 (\cite{Abazajian2009}). 
The update version of SSPP of DR8 (\cite{Lee2008}) provides $T_{\rm eff}=6402\,\mathrm{K}$. 
Our derived $T_{\rm eff}$ of 6088 K is slightly lower than the value of \citet{Aoki2013}, but is within the uncertainty of these estimates.
 
\begin{figure}[htbp]
  \begin{center}
  \FigureFile(80mm,57.8mm){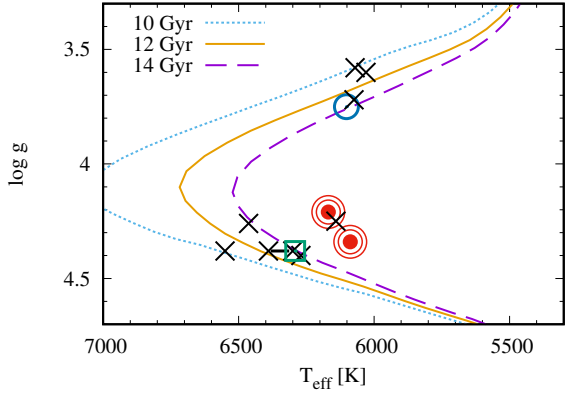}
  \end{center}
  \caption{Location of observed stars in HR diagram. Red dots with two open circles show two CEMP-no stars in this study. LAMOST J1305+2815 is shown with blue open circle. Stellar parameters of G64-12 from previous studies are shown with black crosses and those from our analysis is the open green square. \citet{Aoki2006} adopted $T_{\rm eff}=6390\,\mathrm{K}$ for G64-12, though their Balmer line analysis derive $T_{\rm eff}=6300\,\mathrm{K}$. Both of them are shown in the figure and connected to each other by the black solid line. $Y^2$ isochrones \citep{Demarque2004} assuming $[\mathrm{Fe/H}]=-3$ and $T=10,\,12,\,14,\,\mathrm{Gyr}$ are also shown.}
  \label{figHR}
\end{figure}

\begin{table*}[tpbh]
\tbl{Comparison of stellar parameters and abundances of G64-12.}{
\begin{tabular}{llrrrrrrrrr}\hline
&&This work&\multicolumn{8}{c}{Previous studies\footnotemark[$*$]}\\ \cline{4-11}
                                                                        &       &         & (1)   & (2)   & (3)                        & (4)   & (5)                 & (6)      & (7)   & (8)     \\ \hline
                                  $T_{\rm eff}$\footnotemark[$\dagger$] &       & 6291    & 6270  & 6390  & 6141                       & 6550  & 6074                & 6463     & 6070  & 6030    \\
                                                                        &       & B       & B     & Hd\& C\footnotemark[$\ddagger$] & C     & C              & E     & IRFM     & E     & E       \\
                                  $\log g$\footnotemark[$\dagger$]      &       & 4.38    & 4.4   & 4.38  & 4.25                       & 4.68  & 3.72                & 4.26     & 3.58  & 3.6     \\
                                                                        &       & B       & Ic    & Ic    & I                          & I     & I\footnotemark[$\$$]& P        & I     & Ic      \\
                                  $v_{\rm turb}$                        &       & 1.5     & 1.5   & 1.6   & 1.49                       & 1.9   & 1.19                & 1.65     & 1.46  & 1.2     \\
                                  $[\mathrm{Fe/H}]$                     & Fe I  & -3.36   & -3.37 & -3.2  & -3.44                      & -3.24 & -3.47               & -3.29    & -3.56 & -3.58   \\
                                  $A$(Li)                               & Li I  & 2.22    & 2.18  & 2.3   &                            &       &                     & 2.36     &       & 2.08    \\
                                  $[\mathrm{C/Fe}]$                     & CH    & $<$0.93 &       & 0.49  & 0.71                       &       &                     & 1.07     &       & $<$0.91 \\
                                  $[\mathrm{Na/Fe}]$                    & Na I  & -0.09   & -0.06 & -1.13 &                            &       &                     & -0.09    &       & -0.06   \\
                                  $[\mathrm{Mg/Fe}]$                    & Mg I  & 0.36    & 0.37  & 0.46  & 0.55                       & 0.32  & 0.63                & 0.48     & 0.76  & 0.50    \\
                                  $[\mathrm{Ca/Fe}]$                    & Ca I  & 0.51    &       & 0.46  & 0.48                       & 0.56  & 0.61                & 0.50     & 0.51  & 0.62    \\
                                  $[\mathrm{Sc/Fe}]$                    & Sc II & 0.23    &       &       &                            &       &                     & 0.17     &       & -0.24   \\
                                  $[\mathrm{Ti/Fe}]$                    & Ti I  & 0.66    &       &       &                            &       & 0.53                & 0.74     &       & 0.70    \\
                                  $[\mathrm{Ti/Fe}]$                    & Ti II & 0.57    &       & 0.38  &                            & 0.64  & 0.47                &    0.52  & 0.60  & 0.11    \\
                                  $[\mathrm{Cr/Fe}]$                    & Cr I  & -0.11   &       & -0.18 &                            & -0.10 & 0.07                & -0.09    & -0.10 & -0.20   \\
                                  $[\mathrm{Fe/Fe}]$                    & Fe II & 0.11    &       & 0.13  &                            & -0.02 & 0.07                & 0.05     & 0.00  & 0.19    \\
                                  $[\mathrm{Sr/Fe}]$                    & Sr II &         & 0.08  & 0.18  & 0.06                       & 0.18  &                     & 0.07     &       & -0.10   \\
                                  $[\mathrm{Ba/Fe}]$                    & Ba II & -0.08   &       & -0.25 & 0.07                       & 0.06  & -0.2                & -0.07    & -0.25 & -0.46   \\ \hline
\end{tabular}}
\label{tabg6412}
\begin{tabnote}

\noindent\footnotemark[$*$] (1): \citet{Aoki2009}, (2): \citet{Aoki2006}, (3): \citet{Barklem2005}, (4): \citet{Ishigaki2012}, (5): Stephens and Boesgaard (2002), (6): \citet{Placco2016}, (7): \citet{Ishigaki2010}, (8): \citet{Roederer2014}.

\noindent\footnotemark[$\dagger$] The methods used to determine $T_{\rm eff}$ and $\log g$ are shown. B: Balmer line analysis, C: color, IRFM: Infrared flux method, Ic: Isochrones, E: excitation equilibrium of Fe~I, I: ionization equilibrium of Fe~I and Fe~II, Hd: H$\delta$ index, P: photometric indices.

\noindent\footnotemark[$\ddagger$] \citet{Aoki2006} determines $T_{\rm eff}$ from several methods. Although their Balmer line analysis yields $T_{\rm eff}=6300\,\mathrm{K}$, they adopt the average of $T_{\rm eff}$ from colors and from H$\delta$ index.

\noindent\footnotemark[$\$$] Their $\log g$ is the averaged value of one which achieve ionization equilibrium of iron and one which achieve that of titanium.
\end{tabnote}
\end{table*}
\subsection{Abundance analysis}
We apply standard analysis to equivalent widths measured for clean lines that are apparently not affected by blending after each line is fitted with a Gaussian profile. 
The line list is taken from \citet{Aoki2013} except for the CH~G band and Li~I doublet.
The effect of hyperfine splitting of Ba~II and Sc~II are included following \citet{McWilliam1998} and \citet{McWilliam1995}, respectively.
Uncertainty of determined abundance is estimated as $\sqrt{\sigma^2/N}$, where $\sigma$ is the standard deviation of abundances determined from individual lines, and $N$ is the number of lines measured for the element. 
For elements with less than 3 measurable absorption lines, the uncertainty of the abundance measurement is estimated by adopting the $\sigma$ value of Fe abundances from Fe~I lines in the above formula. 
We also include errors coming from uncertainty in stellar parameters. 
The errors are estimated by measuring abundances using model atmosphere with the offset in stellar parameters, corresponding to the uncertainty in derived stellar parameters.

Abundances of carbon and lithium are determined by fitting observed spectra with synthetic spectra. 
Carbon abundance is determined from the CH~G band around $4324\,\mathrm{\AA}$ and the lithium abundance is from the doublet around $6708\,\mathrm{\AA}$. 
Synthetic spectra are convolved with instrumental profile, rotational profile and/or macroturbulance using a Gaussian profile of which FWHM is estimated from iron lines. 
The line list of CH~G band is taken from \citet{Masseron2014}, and that of Li~I doublet is from the line list of $^7\mathrm{Li}$ in \citet{Smith1998}. 
Sometimes observed spectrum is shifted vertically to derive appropriate continuum level and horizontally to match the line center. 
Abundance is determined as the value which gives the minimal $\chi^2$. 
We confirm that the best-fit synthesized spectra well reproduce the observed one, in order to exclude incorrect fitting due to bad columns and/or cosmic rays. 
The error in fitting estimated with a Monte-Carlo method is $\sim 0.02\,\mathrm{dex}$, which is much smaller than $\sigma$ of Fe~I lines. 
We adopt the $\sigma$ value of Fe lines as the uncertainty of abundances of carbon and lithium in order to take into account other sources of error, such as uncertainty in oscillator strength or normalization.
We also include errors that stem from uncertainties in stellar parameters which are estimated by the same way as for the other elements.

Results of abundance analysis are given in Table \ref{tabab}. 
Internal errors are included, but systematic errors due to determination methods of stellar parameters are not.

We compare derived abundances of G64-12 with previous studies in Table \ref{tabg6412}. 
In particular, abundances of $[\mathrm{Fe/H}]$, $A(\mathrm{Li})$ and $[\mathrm{Ba/Fe}]$ of our study are consistent with previous studies which derived similar stellar parameters.
We note that, although \citet{Placco2016} have determined the carbon abundance of G64-12, the strongest part of CH~G band of our spectra unfortunately hits bad columns.
The difference between the carbon abundance measured by \citet{Placco2016} and our upper limit, is likely due to the different choice of effective temperature.

Fitting results for CH~G band of LAMOST J1410-0555 and SDSS J1424+5615 are shown in Figure \ref{figc}. 
For SDSS J1424+5615, there is no detectable Ba~II lines in its spectrum. 
Figure \ref{figba} shows the spectrum around the strongest Ba~II line and the model spectrum assuming $[\mathrm{Ba/Fe}]=-0.67$. 
We have also examined a presence of Eu II line at $\sim 4129\,\mathrm{\AA}$, but do not find any absorption feature at this wavelength ($[\mathrm{Eu/Fe}]<+2.0$).
Since both stars have $[\mathrm{C/Fe}]>1.0$ and $[\mathrm{Ba/Fe}]<0$, they are unquestionably CEMP-no stars.
\begin{figure}[htbp]
  \begin{center}
  \FigureFile(80mm,100mm){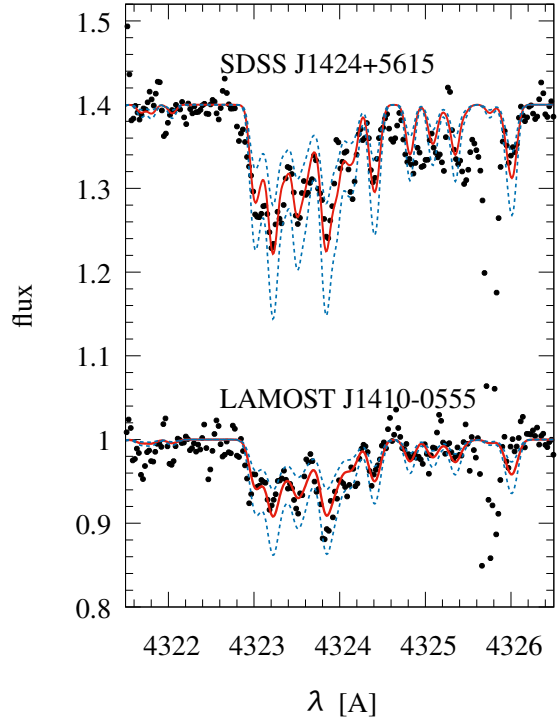}
  \end{center}
  \caption{Fitting results for CH~G bands around $4322\,\mathrm{\AA}$. The spectrum of SDSS J1424+5615 is shifted vertically. The red solid lines show the best fit spectra and the blue dashed ones show the spectra with offset of $\pm 0.2\,\mathrm{dex}$ in $[\mathrm{C/Fe}]$.}
  \label{figc}
\end{figure}
\begin{figure}[htbp]
  \begin{center}
  \FigureFile(80mm,44.4mm){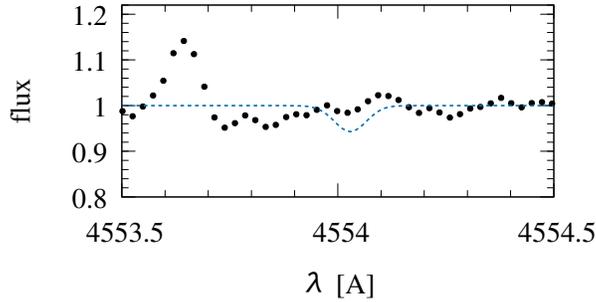}
  \end{center}
  \caption{Comparison of a model spectrum ($[\mathrm{Ba/Fe}]=-0.67$) and SDSS J1424+5615 spectrum for Ba~II $4554\,\mathrm{\AA}$. This results in upper limit of Ba abundance $[\mathrm{Ba/Fe}]<-0.67$ for this object.}
  \label{figba}
\end{figure}

Fitting results for the Li~I doublet of LAMOST J1410-0555 and SDSS J1424+5615 are presented in Figure \ref{figli}. 
Lithium abundances of both stars are $A({\rm Li})\sim 2.15\,\mathrm{dex}$, placing them on the Spite plateau.
\begin{figure}[htbp]
  \begin{center}
  \FigureFile(80mm,100mm){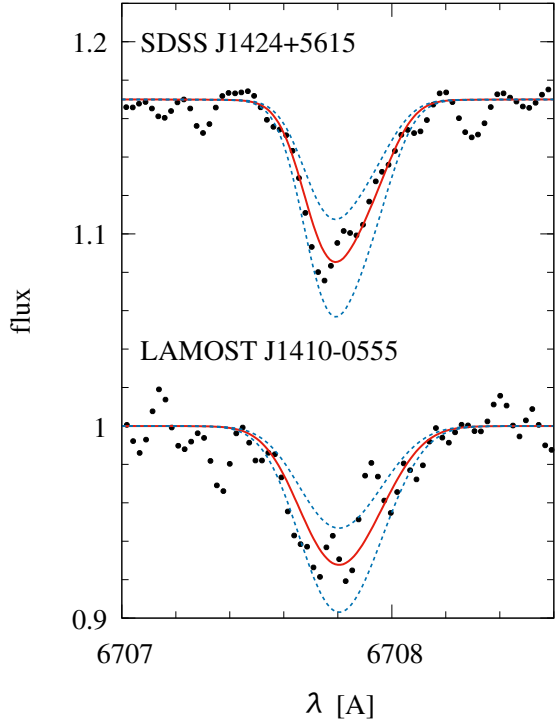}
  \end{center}
  \caption{Same as Figure \ref{figc} but for Li~I 6708 $\mathrm{\AA}$.}
  \label{figli}
\end{figure}

It should be noted that both of the two CEMP-no stars show large enhancement in magnesium abundance, while other $\alpha$ elements show no excess. 
Sodium is also enhanced in these stars. 
Although these abundance anomalies may be a clue to the understanding of CEMP-no stars, it is beyond the scope of this paper.

\section{Discussion}\label{secdis}
\subsection{Data compilation}
\begin{figure*}
  \begin{center}
  \FigureFile(160mm,112mm){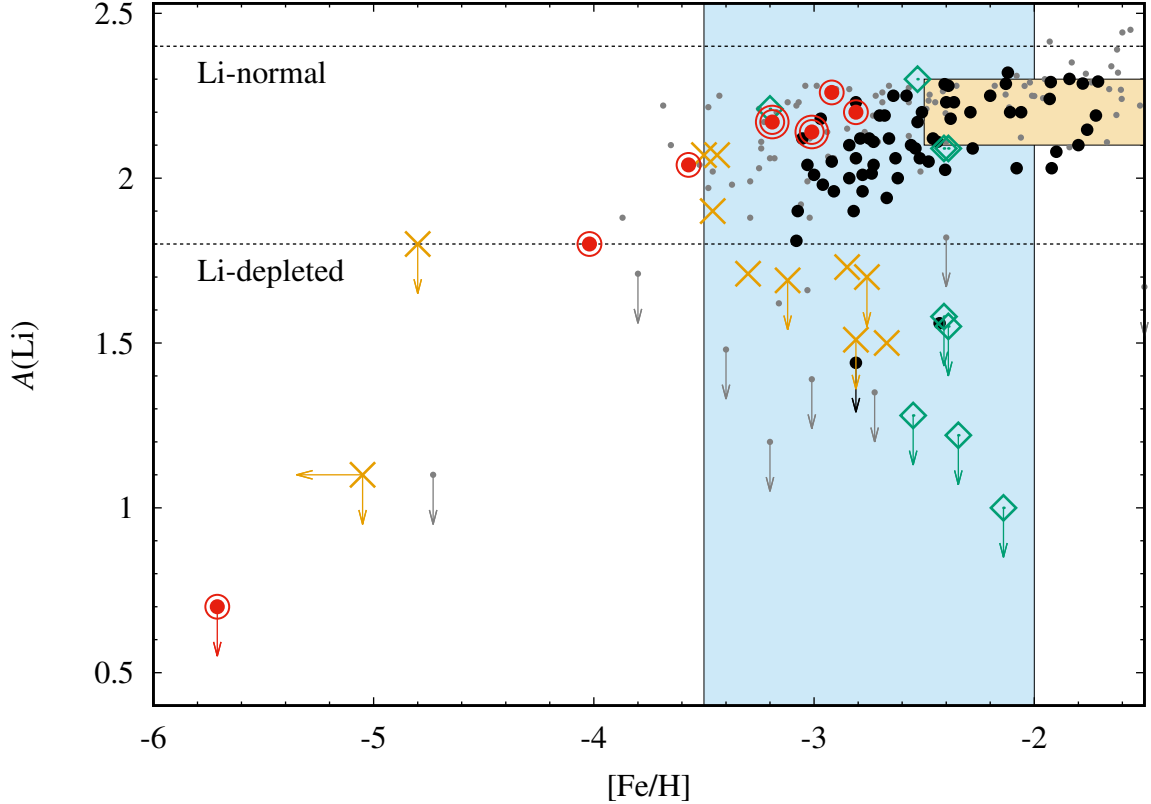}
  \end{center}
  \caption{$A(\mathrm{Li})$ as a function of $[{\rm Fe / H}]$. Black dots show carbon-normal stars, grey dots are stars without carbon measurement and CEMP candidates, green squares are CEMP-s stars, red dots with a circle are CEMP-no stars and yellow crosses are CEMP stars that can not be classified. Our new CEMP-no stars are red dots with double circles. Spite plateau ($A(\mathrm{Li})=2.2\pm 0.1;\,[\mathrm{Fe/H}]>-2.5$) is indicated by the yellow region. Note that lithium abundance of CEMP-no stars is comparable to that of carbon-normal stars at $-3.5<[\mathrm{Fe/H}]<-2.0$ (the blue shaded region), whereas that of CEMP-s stars is lower.}
  \label{figlipl}
\end{figure*}
Figure \ref{figlipl} shows lithium abundances as a function of $[\mathrm{Fe/H}]$ for our targets together with the data taken from the SAGA database (\cite{Suda2008}; \cite{Suda2011}; \cite{Yamada2013}). 
The database compiles abundances of halo stars from various literature. 
It sometimes contains multiple measurements for one star if more than one paper provide abundance. 
Followings are the criteria we adopt to select main-sequence turn-off metal-poor stars and to classify stars into subclasses.
\begin{itemize}
\item {\it Main-sequence turn-off metal-poor stars}: All measurements by different studies give i) $T_{\rm eff}\geq5800\,\mathrm{K}$, ii) $\log g\geq3.7$, and iii) $[\mathrm{Fe/H}]\leq-1.5$. 
\item {\it C-normal/CEMP stars}: All measurements give $[\mathrm{C/Fe}]$ below/above 0.7. When at least one of the measurements gives $[{\rm C/Fe}]>0.7$ and the others give $[{\rm C/Fe}]<0.7$ for the same star, it is classified into ``CEMP candidates''.
\item {\it CEMP-no/CEMP-s stars}: Simple cut in $[\mathrm{Ba/Fe}]$ is traditionally used to classify CEMP stars. Most of UMP stars, however, do not have tight upper limit on $[\mathrm{Ba/Fe}]$ enough to classify them into CEMP-no stars because of their very low iron abundances. Taking into consideration the increasing trend in $[\mathrm{Ba/Fe}]$ with $[\mathrm{C/Fe}]$ among CEMP-s stars (see e.g., \cite{Aoki2007}) , we introduce new criteria for CEMP-s/CEMP-no stars: For $[\mathrm{C/Fe}]<2$, objects with $[\mathrm{Ba/Fe}]>1$ are classified into CEMP-s and those with $[\mathrm{Ba/Fe}]<0$ are CEMP-no stars. For $[\mathrm{C/Fe}]>2$, $[\mathrm{Ba/C}]>-1$ and $[\mathrm{Ba/C}]<-2$ are adopted as criteria for CEMP-s and CEMP-no stars, respectively (see Figure \ref{figbac}). Eu abundance is not used to classify CEMP stars, thus CEMP-s stars could contain some CEMP-r/s stars. In the case that no measurement of barium abundance is available, or its barium abundance falls between the borders, the star is dealt with as ``Unclassifiable CEMP stars'' in this paper. 
\end{itemize}

If a CEMP star has multiple measurements, we select one of them (the first reference for each star in the Table \ref{cemp_tab}) as the abundance of the star to plot in Figure \ref{figlipl}, \ref{figbac} and \ref{figtli}. 
For carbon normal stars and ``CEMP candidates'', we simply adopt the average of multiple measurements in our discussion.

Our new criteria makes it possible to classify UMP stars based on their current upper limits on barium abundances. In our literature sample, our new criteria enables to classify HE1327-2326 ($[\mathrm{C/Fe}]=4.06$ and $[\mathrm{Ba/Fe}]<1.24$; \cite{Frebel2008}) into a CEMP-no star.

In addition, an EMP star, SDSS J0212+0137 \citep{Bonifacio2015}, is classified into a CEMP-no star, though its barium abundance ($[\mathrm{Ba/Fe}]=+0.02$) is slightly higher than the traditional criteria of CEMP-no stars.
It should be noted that, though one star in the literature sample, HE0024-2523 ($[\mathrm{Fe/H}]=-2.72$; \cite{Lucatello2003}) was classified into a CEMP-s star with the traditional criteria, it can not be classified into a CEMP-s nor CEMP-no star with the new criteria.

Recently, \citet{Yoon2016} argued that CEMP stars should be classified according to $A(\mathrm{C})$ as $A(\mathrm{C})>7.1$ to be CEMP-s and $A(\mathrm{C})\leq 7.1$ to be CEMP-no stars. 
Although most of the CEMP-no and CEMP-s stars in Table \ref{cemp_tab} are not affected by the choice of the classification, we also examine the lithium abundances for the various stellar sub-classes using this criteria.

\begin{longtable}{lrrrrrrrr}
\caption{Abundances of CEMP stars. First reference for each star in this table is adopted in this study. When the first one does not provide some of the parameters, the second one is referred for those parameters.}\label{cemp_tab}
\hline
Object & $T_{\rm eff}$ & $\log g$ & $[\mathrm{Fe/H}]$ & $A({\rm Li})$ & $[\mathrm{C/Fe}]$ & $[\mathrm{Ba/Fe}]$& $A(\mathrm{C})$\footnotemark[$\dagger$] & Reference  \\ \hline\hline
\endfirsthead
\hline
\endhead
\hline
\endfoot
\hline
\multicolumn{9}{p{80mm}}{
\parbox{140mm}{\vspace{2mm}1: \citet{Roederer2014}, 2: \citet{Sbordone2010}, 3: \citet{Frebel2008}, 4: \citet{Aoki2006}, 5: \citet{Frebel2006}, 6: \citet{Li2015a}, 7: \citet{Bonifacio2015}, 8: \citet{Johnson2007}, 9: \citet{Preston2001}, 10: \citet{Masseron2012},11 \citet{Thompson2008}, 12: \citet{Aoki2008}, 13: \citet{Aoki2002}, 14: \citet{Norris1997}, 15: \citet{Charbonnel2005}, 16: \citet{Behara2010}, 17: \citet{Aoki2013}, 18: \citet{Bonifacio2007}, 19: \citet{Bonifacio2009}, 20: \citet{Sivarani2006}, 21: \citet{Lucatello2003}, 22: \citet{Aoki2009}, 23: \citet{Barklem2005}, 24: \citet{Sneden2003}, 25: \citet{Lai2008}, 26: \citet{Placco2016}, 27: \citet{Akerman2004} \\
\footnotemark[$\dagger$] $A(\mathrm{C})$ is converted using the solar abundance of \citet{Asplund2009}, though other values are directly from SAGA database, i.e. there remains solar abundance difference.\\
\footnotemark[$\ddagger$] See Table \ref{tabg6412} for the results by other studies.
}
}\\
\endlastfoot
\multicolumn{9}{c}{CEMP-no stars}\\\hline
LAMOST J1410-0555& 6169 & 4.21 & -3.19 & 2.17    & 1.53 & -0.33   & 6.77 & this work\\
SDSS J1424+5615  & 6088 & 4.34 & -3.01 & 2.14    & 1.49 & $<$-0.69 & 6.91 & this work\\
CS29493-050      & 6270 & 3.8  & -2.89 & 2.26    & 0.81 & -0.72   & 6.35  & 1\\
CS29514-007      & 6400 & 3.85 & -2.81 & 2.2     & 0.88 & -0.16   & 6.50 & 1\\
                 & 6281 & 4.1  & -2.8  & 2.24    &      &         &  & 2\\
HE1327-2326      & 6180 & 3.7  & -5.71 & $<$0.70 & 4.06 & $<$1.24 & 6.78 & 3\\
                 & 6180 & 3.7  & -5.66 & $<$1.5  & 4.26 & $<$1.46 & 7.03 & 4\\
                 & 6180 & 3.7  & -5.7  &         & 4.0  &         &  & 5\\
LAMOSTJ1253+0753 & 6030 & 3.7  & -4.02 & 1.80    & 1.59 & $<$-0.30& 6.00 & 6\\
SDSSJ0212+0137   & 6333 & 4.0  & -3.57 & 2.04    & 2.21 & 0.02    & 7.07 & 7\\\hline

\multicolumn{9}{c}{CEMP-s stars}\\\hline
CS22879-029    & 5920 & 3.7  & -2.55 & $<$1.28 & 1.26 & 1.18& 7.10& 1\\
               & 6300 & 3.90 & -1.93 &         & 1.30 &     & 7.80& 8\\
CS22881-036    & 5940 & 3.7  & -2.34 & $<$1.22 & 1.99 & 2.03& 8.08& 1\\
               & 6200 & 4.0  & -2.05 &         & 1.96 & 1.93& 8.34& 9\\
CS22896-136    & 6190 & 3.85 & -2.41 & $<$1.58 & 1.08 & 1.22& 7.10& 1\\
CS22956-102    & 6220 & 3.85 & -2.39 & $<$1.55 & 2.03 & 1.81& 8.07& 1\\
CS22949-008a   & 6300 & 4.0  & -2.14 & $<$1.0  & 1.44 & 1.41& 7.73& 10\\
CS22964-161A   & 6050 & 3.7  & -2.39 & 2.09    & 1.35 & 1.45& 7.39& 11\\
CS22964-161B   & 5850 & 4.1  & -2.41 & 2.09    & 1.15 & 1.37& 7.17& 11\\
CS31062-012    & 6200 & 4.3  & -2.53 & 2.3     & 2.14 & 2.08& 8.04& 12\\
               & 6250 & 4.5  & -2.55 &         & 2.1  & 1.98& 7.98& 13\\
               & 6000 & 3.8  & -2.74 &         & 2.15 & 2.01& 7.94& 14\\
               & 5901 & 4.5  & -3.50 & 1.973   &      &     & & 15\\
SDSSJ1036+1212 & 6000 & 4.0  & -3.2  & 2.21    & 1.47 & 1.17& 6.70& 16\\
               & 5850 & 4.0  & -3.47 &         & 1.84 & 1.35& 6.80& 17\\\hline
\multicolumn{9}{c}{Unclassifiable CEMP stars}\\\hline
CS22882-012 & 6290 & 3.8  & -2.76 & $<$1.7  & 1.09 & 0.61 &6.76  & 1\\
CS22945-017 & 6080 & 3.7  & -2.89 & $<$1.51 & 1.78 & 0.49 &7.32  & 1\\
CS29504-006 & 6150 & 3.7  & -3.12 & $<$1.69 & 1.39 & 0.21 &6.70  & 1\\
SDSSJ1742+2531   & 6345 & 4.0  & -4.80  & $<$1.80 & 3.63 & $<$1.72 & 7.26 & 7\\
SDSSJ1035+0641   & 6262 & 4.0  &$<$-5.05& $<$1.1  & 3.08 & $<$2.25 & 6.70 & 7\\
CS29527-015 & 6242 & 4.00 & -3.44 & 2.07    &      &      &  & 18\\
            & 6242 & 4.00 & -3.55 &         & 1.18 &      &6.06  & 19\\
CS29528-041 & 6150 & 4.00 & -3.3  & 1.71    & 1.59 & 0.97 &6.72  & 20\\
CS31080-095 & 6050 & 4.50 & -2.85 & 1.73    & 2.69 & 0.77 &8.27  & 20\\
HE0024-2523 & 6625 & 4.3  & -2.72 & 1.5     & 2.6  & 1.46 &8.31  & 21\\
HE1148-0037 & 5990 & 3.7  & -3.46 & 1.90    &      &      &  & 22\\
            & 5964 & 4.16 & -3.47 &         & 0.84 &      &5.80  & 23\\
HE1413-1954 & 6302 & 3.80 & -3.50 & 2.04    &      &      &  & 2\\
            & 6533 & 4.59 & -3.19 &         & 1.45 &      &6.69  & 23\\\hline
\multicolumn{9}{c}{CEMP candidates}\\\hline
CS22953-037 & 6364 & 4.25 & -2.89 & 2.16 & 0.37    &          &5.91  & 19\\
            & 6150 & 3.7  & -3.21 & 1.97 & 0.84    & $<$-0.67 &6.06  & 1\\
CS22964-214 & 6800 & 4.5  & -2.30 &      & 1.0     & -0.12    &7.13  & 24\\
            & 6180 & 3.75 & -2.95 & 2.02 & 0.62    & -0.54    &6.10  & 1\\
CS22965-054 & 6205 & 3.73 & -3.09 &      & $<$1.07 & $<$-0.48 &  & 25\\
            & 6310 & 3.9  & -2.84 & 2.16 &         &          &  & 22\\
            & 6089 & 3.75 & -3.04 &      & 0.62    &          &6.01  & 19\\
            & 6050 & 3.7  & -3.17 & 2.16 & 0.74    & $<$-0.45 &6.00  & 1\\
G64-12      & 6462 & 4.26 & -3.29 & 2.36 & 1.07    & -0.07    &6.21  & 26\\
G64-37      & 6570 & 4.40 & -3.11 & 2.25 & 1.12    & -0.36    &6.44  & 26\\
            & 6318 & 4.16 & -3.12 &      & 0.29    &          &5.72  & 27\\
            & 6396 & 3.85 & -2.96 & 2.08 &         &          &      & 15\\\hline
\end{longtable}
\begin{figure*}
  \begin{center}
  \FigureFile(160mm,112mm){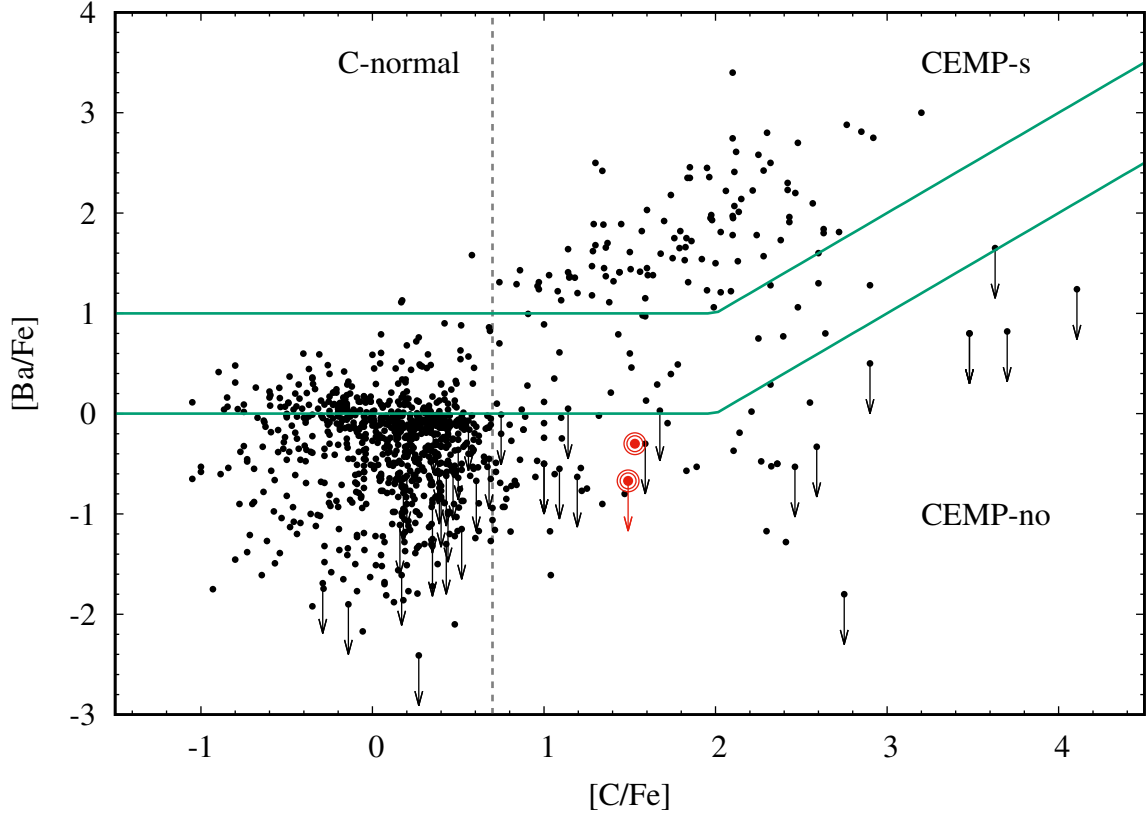}
  \end{center}
  \caption{$[\mathrm{Ba/C}]$ as a function of $[\mathrm{C/Fe}]$. Two solid lines show the criteria used in this study to classify CEMP stars into subclasses. Our new two stars meet the criteria of CEMP-no stars (red dots with two circles). }
  \label{figbac}
\end{figure*}

\subsection{Lithium abundances for CEMP subclasses}
Since the purpose of this paper is to examine lithium abundance in CEMP-no stars with $[{\rm Fe / H}]\sim -3$, the following discussion is focused on stars within the metallicity range of $-3.5<[{\rm Fe / H}]<-2.0$ (blue-shaded region in Figure \ref{figlipl}).

By taking a glance at the figure, the plateau of the lithium abundance can be easily found for carbon-normal stars, though some scatter exists, in particular, at $[{\rm Fe/H}]\lesssim -3$. Most stars without carbon excess marked with black points in the figure have lithium abundance between $A(\mathrm{Li})=1.8$ and $A(\mathrm{Li})=2.4$. 
Thus, we define stars with $1.8<A(\mathrm{Li})<2.4$ as lithium-normal and those with $A(\mathrm{Li})<1.8$ as lithium-depleted. No objects with $A(\mathrm{Li})>2.4$ is found in the present sample. Based on the criteria, only two out of the 63 carbon-normal stars are classified as lithium-depleted stars (Table \ref{tablifrac}). 
One is HE0411-3558 ($T_{\rm eff}=6300\,\mathrm{K}$, $\log g=3.7$, $[\mathrm{Fe/H}]=-2.8$, $[\mathrm{C/Fe}]<0.7$, and $A(\mathrm{Li})<1.44$; \cite{Hansen2015}) and the other is CS22957-019 ($T_{\rm eff}=6070\,\mathrm{K}$, $\log g=3.75$, $[\mathrm{Fe/H}]=-2.43$, $[\mathrm{C/Fe}]=0.34$, and $A(\mathrm{Li})=1.56$; \cite{Roederer2014}). 
\citet{Ryan2002} investigated lithium-depleted stars at higher metallicity ($[\mathrm{Fe/H}]\sim -1.5$) and concluded that the low lithium abundances could be the result of mass transfer. 
Although HE0411-3558 and CS22957-019 show no radial velocity variation as a result of multiple observations, the radial velocity monitoring could be insufficient or the system could be face on.
However, no significant abundance anomalies which can be considered as a result of mass transfer is found within the measurement errors (\cite{Hansen2015} and \cite{Roederer2014}).
 
On the other hand, five of the nine CEMP-s stars are lithium-depleted (green open squares, see also Table \ref{tablifrac}). This can be interpreted as a result of mass transfer from former AGB companions which should have become white dwarfs as discussed in Section \ref{secint}. Other four CEMP-s stars are lithium-normal, this could be the result of small amounts of transferred mass which would lead to carbon- and barium-excesses without severe lithium-depletion (\cite{Masseron2012}, but see also \cite{Stancliffe2010}).

It has been difficult to make any discussion on lithium abundance in CEMP-no stars at $[\mathrm{Fe/H}]\sim -3$ because of the small sample size. We add two CEMP-no stars with lithium measurement to the sample. Together with CS29493-050 and CS29514-007 from \citet{Roederer2014}, it now becomes possible to discuss the lithium abundance in CEMP-no stars for the first time. The four CEMP-no stars with measured lithium abundance in the metallicity range, $-3.5<[{\rm Fe / H}]<-2.0$, are lithium-normal (Table \ref{tablifrac}). 
Furthermore, the two CEMP-no stars at lower metallicity, SDSS J0212+0137 ($[{\rm Fe/H}]=-3.57$ and $A({\rm Li})=2.04$; \cite{Bonifacio2015}) and LAMOST J1253+0753 ($[{\rm Fe/H}]=-4.02$ and $A({\rm Li})=1.8$; \cite{Li2015}) also have normal lithium abundances in our definition. 
Plus very recently, \citet{Placco2016} reported the carbon abundances of the bright EMP stars G64-12 and G64-37 with $[\mathrm{Fe/H}]<-3$ are as high as $[\mathrm{C/Fe}]\sim +1.0$, indicating that these stars are CEMP-no lithium normal stars.
If these two stars are included in the sample of CEMP-no stars, our result that CEMP-no have normal lithium abundances is strengthened.

Fisher's exact test of independence rejects the hypothesis that carbon-normal stars and CEMP-s stars with $-3.5<[\mathrm{Fe/H}]<-2.0$ have the same lithium depleted fraction ($p=0.00017$). 
By contrast, the lithium depleted fraction of CEMP-no stars is not distinguishable from carbon normal stars ($p=1.0$). Although the hypothesis that CEMP-no stars and CEMP-s stars have the same lithium depleted fraction is not rejected ($p=0.11$), CEMP-no stars are more likely to behave like carbon normal stars.  
If we include G64-12 and G64-37 in the CEMP-no sample, the latter $p$-value drops to $0.04$.
This result suggests that the lithium abundances for the majority of the CEMP-no stars are similar to non-carbon enhanced stars contrary to the case of CEMP-s stars.
This seems to be related to the reported difference in binarity of CEMP-no stars and CEMP-s stars (\cite{Starkenburg2014}; \cite{Hansen2016b}; \cite{Hansen2016}), and supports that no significant lithium-depletion is found in $[\mathrm{Fe/H}]\gtrsim -4$ regardless of carbon abundance, excluding CEMP-s stars. 
\begin{table}
  \tbl{Lithium depleted fraction for each subclass.}{
  \begin{tabular}{lrrr}\hline
      & C-normal & CEMP-s & CEMP-no \\ \hline
    Li-normal ($1.8<A(\mathrm{Li}) < 2.4)$ & 61 & 4 & 4 \\ 
    Li-depleted ($A(\mathrm{Li} < 1.8$) & 2 & 5 & 0 \\
    Li-depleted fraction & 0.032 & 0.556 & 0.000 \\ \hline
  \end{tabular}}
  \label{tablifrac}
\end{table}

We also examine the lithium abundance difference between the two CEMP classes using the criterion based on $A(\mathrm{C})$ proposed by \citet{Yoon2016}, who separate CEMP-s and CEMP-no stars at $A(\mathrm{C})=7.1$.
We assume $\pm 0.5\,\mathrm{dex}$ margins to avoid contamination as done when $[\mathrm{Ba/Fe}]$ based criterion is adopted.
Thus we regard stars with $A(\mathrm{C})<6.6$ as CEMP-no stars and those with $A(\mathrm{C})>7.6$ as CEMP-s stars.
Four stars in Table \ref{cemp_tab} are classified into CEMP-no and six are into CEMP-s stars.
The number of lithium-normal stars is four for CEMP-no stars and one for CEMP-s stars.
Thus, our results that CEMP-no stars have normal lithium abundances contrary to CEMP-s stars do not depend on the way of the classification.
 
Temperature or metallicity difference between subclasses could cause the lithium abundance variation. Figure \ref{figtli} shows lithium abundance as a function of effective temperature. There is no clear dependence of lithium abundance on temperature as reported in previous studies (e.g., \cite{Aoki2009}; \cite{Sbordone2010}). 
Moreover, temperature difference among the three subclasses is, if any, not large.

Another concern is metallicity difference among subclasses. In particular, it is known that lithium abundance has a decreasing trend with metallicity at $[\mathrm{Fe/H}]<-3$ (e.g., \cite{Sbordone2010}). 
The average metallicity of CEMP-s stars in our sample is higher than that of CEMP-no stars. 
The lower lithium abundance on average in CEMP-s stars than in CEMP-no stars is opposite to the metallicity dependence of lithium abundances  found in carbon-normal stars, indicating that the metallicity difference between the two classes is not the reason for the difference of their lithium abundances.

\begin{figure}
  \begin{center}
  \FigureFile(80mm,56mm){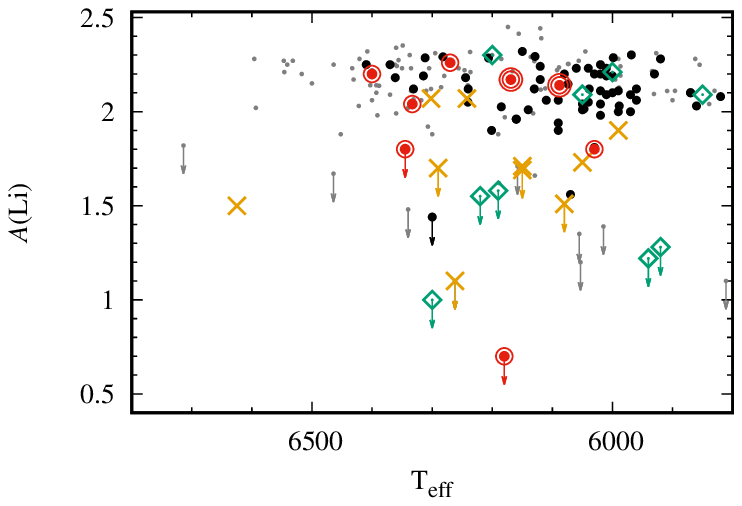}
  \end{center}
  \caption{$A(\mathrm{Li})$ as a function of $T_{\rm eff}$. Symbols are the same as Figure \ref{figlipl}.}
  \label{figtli}
\end{figure}

\subsection{What causes the breakdown of the Spite plateau?}
The results of our measurements, combining those of previous studies, indicate that CEMP-no stars with $[\mathrm{Fe/H}]\sim -3$ have normal lithium abundance as found in carbon-normal stars. 
This suggests that stellar internal structures are not altered so significantly that lithium abundances in CEMP-no stars are lowered during stellar evolution due to the carbon excesses.

Some studies (e.g., \cite{Bonifacio2015}; \cite{Piau2006}) argued a possibility that UMP stars are formed mostly from ejecta from population III stars. If the ejecta from population III stars containing no lithium is  not well-mixed with interstellar medium (ISM), lithium abundance of some fraction of UMP stars could be lowered. 

We investigate this scenario from estimation on dilution factor, defined by
\[
D=M_{\mathrm{ISM}}/M_{\mathrm{ejecta}},
\]
where $M_{\rm ejecta}$ and $M_{\rm ISM}$ are masses of the ejecta and interstellar matter mixed to form UMP stars. In order to reduce lithium abundance from the primordial value ($A(\mathrm{Li})\sim 2.7$), or from the Spite plateau one, to the depleted value ($A(\mathrm{Li})<1.8$), the gas clouds from which UMP stars are formed should be mostly provided by ejecta from population III stars. \textrm{For instance, \citet{Piau2006} estimated $D\sim0.5-1$.} For faint supernovae models, required typical dilution factor is $D> 100$ to reproduce $[{\rm C / H}]$ in CEMP-no stars (e.g., \cite{Takahashi2014}). This dilution factor is too large to make any change in lithium abundance from the ISM value. 
On the other hand, carbon enhancement in spin star models could be related to the low lithium abundance since dilution factor might be small due to gradual expulsion of processed material from the population III stars.

Another explanation for the very low lithium abundances in UMP stars is additional mixing caused by initial high rotational velocity of UMP stars. Numerical simulation shows that low-mass metal-free stars are formed through disk fragmentation around massive protostars (see \cite{Bromm2013} and reference therein). If the formation of UMP stars is a result of disk fragmentation, they should have rotated faster, which results in lithium destruction \citep{Pinsonneault1999}. 
 
Another scenario is the complete lithium destruction at pre-main sequence phase and the lack of late accretion \citep{Fu2015}. Though their model can account for the discrepancy of the observed Spite plateau and the predicted amount of primordial lithium, fine tuning of model parameters is required.

It should be noted that although the current discussion is limited to lithium abundance in unevolved star, two UMP stars, HE0233-0343 ($T_{\rm eff}=6100\,\mathrm{K}$, $\log g=3.4$, $[\mathrm{Fe/H}]=-4.68$ and $A({\rm Li})=1.77$; \cite{Hansen2014}) and SMSS J031300.36-670839.3 ($T_{\rm eff}=5125\,\mathrm{K}$, $\log g=2.3$, $[\mathrm{Fe/H}]<-7.3$, $A({\rm Li})=0.7$; \cite{Keller2014}) have typical lithium abundances in subgiants and red giants, respectively. 
By considering lithium abundance variation along the red giant branch in globular clusters (e.g., \cite{Lind2009a}; \cite{Kirby2016}), their lithium abundances could have been as high as the plateau value when they were main sequence stars. 
\subsection{Constraint on the origin of CEMP-no stars}

Our results indicate that carbon-excess in CEMP-no stars at $[{\rm Fe/H}]\sim-3$ should be brought about through a process without significant lithium depletion. 
This provides a new support for the scenario that CEMP-no stars have different formation mechanism than CEMP-s stars which for the majority have been shown to belong to binary systems and have experienced large amount of mass transfer.
It should be noted that small amount of mass transfer from former AGB companion could cause carbon-excess without barium-excess nor lithium-depletion \citep{Masseron2012}. 
Namely, if the separation of a binary system is large, the amounts of transferred mass would be small.
Such systems can only be detected with long term precise radial velocity monitoring.

Lithium abundances can provide a hint to classify CEMP stars that are not classified by barium abundances (Unclassifiable CEMP stars in Table \ref{cemp_tab}). In particular, three unclassifiable CEMP stars, CS29527-015 (\cite{Bonifacio2007}; \cite{Bonifacio2009}), HE1148-0037 (\cite{Aoki2009}; \cite{Barklem2005}) and HE1413-1954 (\cite{Sbordone2010}; \cite{Barklem2005}) have normal lithium abundances, thus they are likely to be CEMP-no stars, though upper-limit of barium abundance is desirable. 
These three stars also have low $A(\mathrm{C})$, which supports that they are CEMP-no stars, according to \citet{Placco2016}.
It should be noted that, although CS 29527-015 and HE1148-0037 belong to binary systems, both of them are double-lined spectroscopic binary (\cite{Bonifacio2009} and \cite{Sbordone2010}, respectively). 
Therefore it is unlikely that their companions are now white dwarfs having polluted their companion with carbon-rich material during their AGB phase.
On the other hand, the remaining ``unclassifiable'' stars have low lithium abundances, suggesting that they are CEMP-s stars or so-called CEMP {\it low-s} stars \citep{Masseron2010}, though some of them have relatively low $A(\mathrm{C})$ (see Table \ref{cemp_tab}).
Radial velocity variation has been reported only for HE0024-2523 \citep{Lucatello2003}. 
Note that lithium abundances alone can not be a definitive criterion to classify CEMP stars since about a half of CEMP-s stars have normal lithium abundance. In order to classify them by conventional criteria, radial velocity monitoring and measurements of barium abundances are clearly required. 

\section{Summary \& Conclusion}
We have determined lithium abundance in two new unevolved CEMP-no stars at $[{\rm Fe/H}]<-3$ and increased the number of such objects to four. Lithium abundance of all four CEMP-no stars with similar metallicity is not unusual compared to carbon-normal stars, i.e. $1.8<A(\mathrm{Li})<2.2$. These results lead to the following discussions:
\begin{itemize}
\item It is unlikely that low lithium abundance in UMP stars is simply due to high carbon abundance. The low lithium abundance is rather linked to their low iron abundance. 
It is not yet clear what mechanism exists, operating only in iron-poor stars to lower the lithium abundance.
Potential cause of low lithium abundances is initial rapid rotation of UMP stars as a result of disk fragmentation or lack of accretion from lithium-rich ISM after possible complete lithium destruction during pre-main sequence phase. It may also be possible that UMP stars are formed mostly from ejecta of fast rotating population III stars, which does not include lithium. In this case, CEMP-no stars with $[\mathrm{Fe/H}]\sim -3$ are considered to be formed from gas clouds in which ejecta from population III stars are mostly diluted with primordial gas cloud.
\item Lithium abundance in CEMP-no stars behaves differently from that in CEMP-s stars. This is consistent with CEMP-no stars not having experienced large amount of mass transfer like ones CEMP-s stars have, though small amount of mass transfer is possible. This is also consistent with the difference in binarity between these two classes.
\item Lithium abundance can be a hint to classify CEMP stars into subclasses for $[\mathrm{Fe/H}]\gtrsim -3.5$: lithium-normal stars seem to be CEMP-no whereas lithium-poor stars seem to be CEMP-s. However, it can not be a strong criterion since some of CEMP-s stars have normal lithium abundance.
\end{itemize}
In order to clarify the origin of the low lithium abundances in stars with lower metallicity, lithium abundance measurements for many UMP stars are particularly desired. Sample size of UMP stars is expected to increase in the near future as a result of extensive follow-up observations for candidate of metal-poor stars found by e.g., LAMOST or SkyMapper \citep{Keller2007}.

\begin{ack}
We are grateful to Dr. T. C. Beers for useful discussion on CEMP stars. WA and TS were supported by JSPS KAKENHI Grant Numbers JP23224004 and JP16H02168. HNL acknowledge supports by NSFC grants No. 11573032, 11233004, and 11390371.
\end{ack}

\end{document}